\documentclass[twocolumn]{revtex4-2}
\usepackage{graphicx} 
\usepackage{amsmath}

\begin{document}
\title{Polarization-induced Weyl phonons in nonsymmorphic crystals}
\author{Sahal Kaushik}
\affiliation{Nordita, Stockholm University and KTH Royal Institute of Technology, Hannes Alfv\'{e}ns v\"{a}g 12, SE-106 91 Stockholm, Sweden}
\email{sahal.kaushik@su.se}
\begin{abstract}
    In this work, it is shown that in certain nonsymmorphic space groups, electric polarization due to an external electric field or ferroelectric order produces Weyl phonons.  
\end{abstract}
\maketitle

\section{Introduction}
Weyl fermions are Berry monopoles in the electronic dispersion relation \cite{weyl}. They have well-defined chirality, are topologically protected, and in the simplest case, have linear dispersion. They always occur in pairs of opposite chirality \cite{nnt1,nnt2}, and contribute effects such as chiral photocurrents \cite{de2017quantized} and negative longitudinal magnetoresistance \cite{cme}. Because of their topology, a material with Weyl fermions has Fermi arcs at a boundary \cite{weyl}. 

Certain materials also have multifold Weyl fermions, which, like Weyl fermions are Berry monopoles, but of charge greater than $\pm 1$. They are protected by both symmetry and topology; if the symmetry is broken they break into multiple Weyl fermions \cite{bradlyn2016beyond}.

Similar to fermions, phonons can also have Berry monopoles in their dispersion \cite{tiantian}. These Weyl phonons are also topologically protected and have well-defined chirality. Just like Weyl fermions, Weyl phonons are also associated with surface arcs. They have been observed in materials such as FeSi \cite{fesi}.

\section{Creating Weyl Phonons from Perturbations}
Ferroelectric order and/or external magnetic fields are known to produce Weyl fermions in certain materials such as half-Heuslers \cite{magnet1,magnet2}. They can be used to engineer configurations of Weyl cones  \cite{cano2017chiral}. Similarly, ferroelectric order is also known to produce Weyl fermions in certain materials \cite{ferroelectric1,ferroelectric2,ferroelectric3}. We investigate the conditions under which ferroelectric order or polarization induced by external field can produce Weyl phonons.

The chirality $\chi$ of Weyl phonons, is invariant under time reversal $T$ and flips under inversion $P$. Both these symmetries flip crystal momentum Therefore, if the crystal has both symmetries, a left handed phonon coincides with a right handed phonon, and they form a Dirac phonon. In order for a material to have Weyl phonons, it needs to have broken inversion or broken time reversal, or both. Broken inversion can produce sets of $4$ Weyls, while broken time-reversal can produces sets of $2$

We are interested in creating Weyl phonons from a small perturbation. Since electric polarization breaks inversion, it can create sets of 4 Weyls.

Since we are interested in the effects of small perturbation, we focus on effects that are linear, as opposed to quadratic and higher order.

Let us consider a situation where an electric polarization $\Vec{P}$ creates a Weyl node of positive chirality at $\Vec{k}_{1+}(\Vec{P})$. A crystal with opposite polarization has a node of positive chirality at $\Vec{k}_{1+}(-\Vec{P})$. However, these two crystals are related by $PT$, which flips chirality and preserves momentum, so in the crystal with polarization $\Vec{P}$, there is necessarily a Weyl cone of negative chirality at $\Vec{k}_{1-}(\Vec{P}) = \Vec{k}_{1+}(-\Vec{P})$. 

If we take the limit $\Vec{P}\to \Vec{0}$, these nodes merge. They form a Dirac cone at $\Vec{k_1}$, i.e. a fourfold degeneracy (there is also a possibility they annihilate to form a twofold degeneracy, but that is quadratic and not robust against symmetry-preserving perturbations). Since we are considering linear effects, 
\begin{align}\label{eq:inv}
[\Vec{k}_{1-}(\Vec{P})-\Vec{k}_1] = -[\Vec{k}_{1+}(\Vec{P})-\Vec{k}_1] 
\end{align}

Now let us consider the effects of time reversal. Since it preserves chirality but flips momentum, there is also a Weyl cone of positive chirality at $\Vec{k}_{2+}(\Vec{P}) = -\Vec{k}_{1+}(\Vec{P})$ and one of negative chirality at $\Vec{k}_{2-}(\Vec{P}) = -\Vec{k}_{1-}(\Vec{P})$, which merge at $\Vec{k}_2 = -\Vec{k}_1$. Because of (\ref{eq:inv}), $\Vec{k}_1\neq\Vec{k}_2$. Therefore, these vectors must not be at time-reversal invariant points (TRIMs).

\begin{figure}
    \centering
    \includegraphics[scale=0.25]{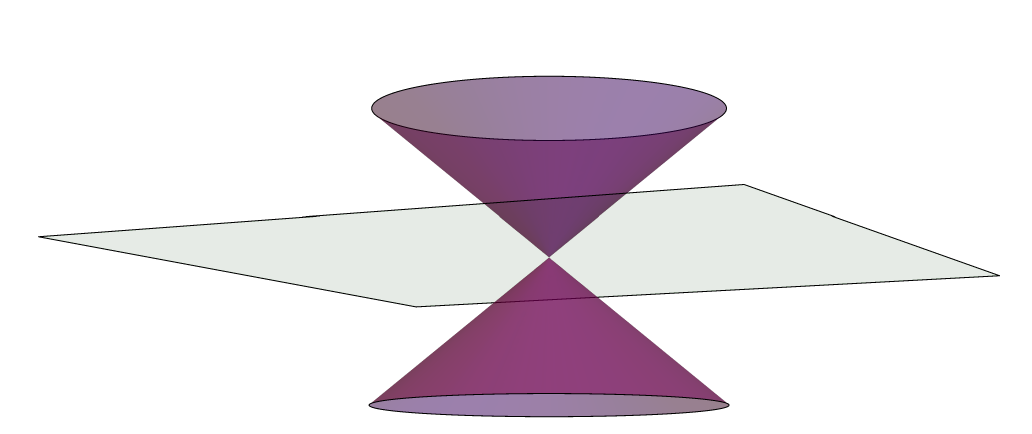}\includegraphics[scale=0.25]{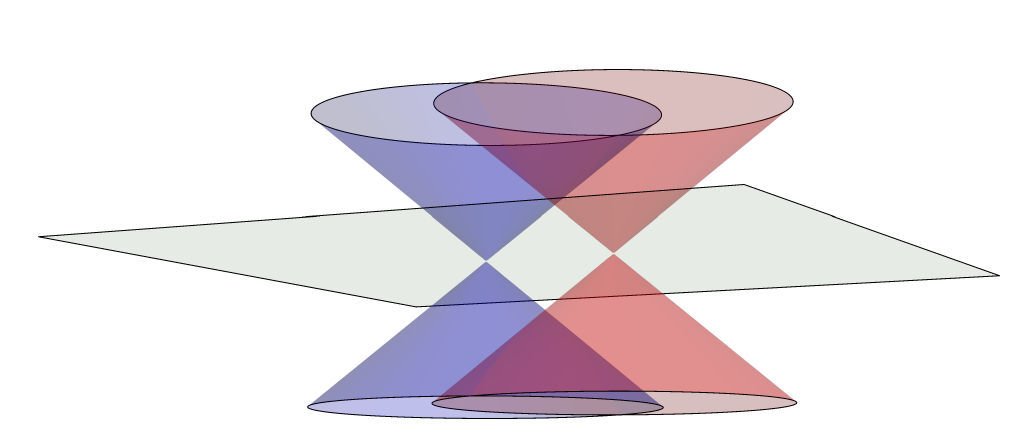}
    \caption{A Dirac cone splitting into two Weyls due to perturbation}
    \label{fig:weyls}
\end{figure}

The Dirac Hamiltonian is of the form 
\begin{equation}
   H = \sum v_{ij}k_i\tau_3\sigma_j
\end{equation}
Where the momentum $k$ is to measured from the Dirac point, the matrices $\sigma_i$ act on the spin and $\tau_i$ on the chirality degrees of freedom. The chirality of the phonons is given by $\gamma_5 = \tau_3$. The Dirac Hamiltonian consists of two copies of the Weyl Hamiltonian 
\begin{equation}
    H = \pm v_{ij}k_i\sigma_j
\end{equation}
If there is a perturbation of the form
\begin{equation}
    \Delta H = a_{ij}P_i\tau_0\sigma_j
\end{equation}
it does not mix states of opposite chirality, but shifts the two Weyl nodes in opposite directions, separating them, as shown in Fig~\ref{fig:weyls}

The Dirac wavefunctions have spin $1/2$, while phonons are bosons. Dirac phonons are possible only in nonsymmporphic lattices, where the fractional phases induced by fractional translations add up to give a $-1$ phase under a full rotation. This is similar to how nonsymmorphic lattices host fermions with integer spin \cite{bradlyn2016beyond}

We have the following requirements for a system where Dirac phonons split into Weyl phonons due to polarization:
\begin{itemize}
    \item There must be Dirac phonons, which are possible in nonsymmorphic space groups at high symmetry points
    \item These high symmetry points must not be TRIMs
    \item These points must \textbf{not} be invariant under residual reflections after symmetry breaking, and cannot transform into each other under residual rotations.
\end{itemize}
 There are four space groups that satisfy all these conditions \cite{bilbao1,bilbao2,bilbao3,bilbaovec1,bilbaovec2}: 
 \begin{enumerate}
     \item $Ibca  (73)$ (body-centered ortorhombic; $\Vec{k} = (\pi,\pi,\pi)$) 
    \item $I4_1/acd (142)$ (body-centered tetragonal; only for polarization along $a$ or $b$; $\Vec{k} = (\pi,\pi,\pi)$)
    \item $Ia\Bar{3} (206)$ (body-centered cubic (BCC); $\Vec{k} = (\pi,\pi,\pi)$)  
    \item $P\Bar{3}c1 (165)$ (hexagonal; $\Vec{k} = (2\pi/3,2\pi/3,\pi)$) 
\end{enumerate}
In the next section, we explicitly show Weyl phonons in toy models.

\section{Toy Models}
\subsection{Cubic}
\begin{figure}
     \centering
     \includegraphics[scale=0.32]{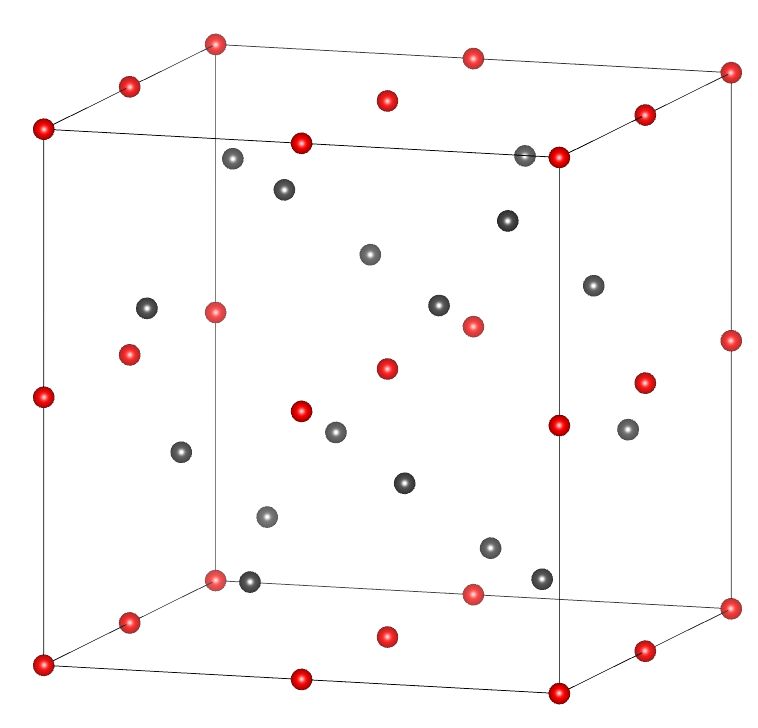}\includegraphics[scale=0.27]{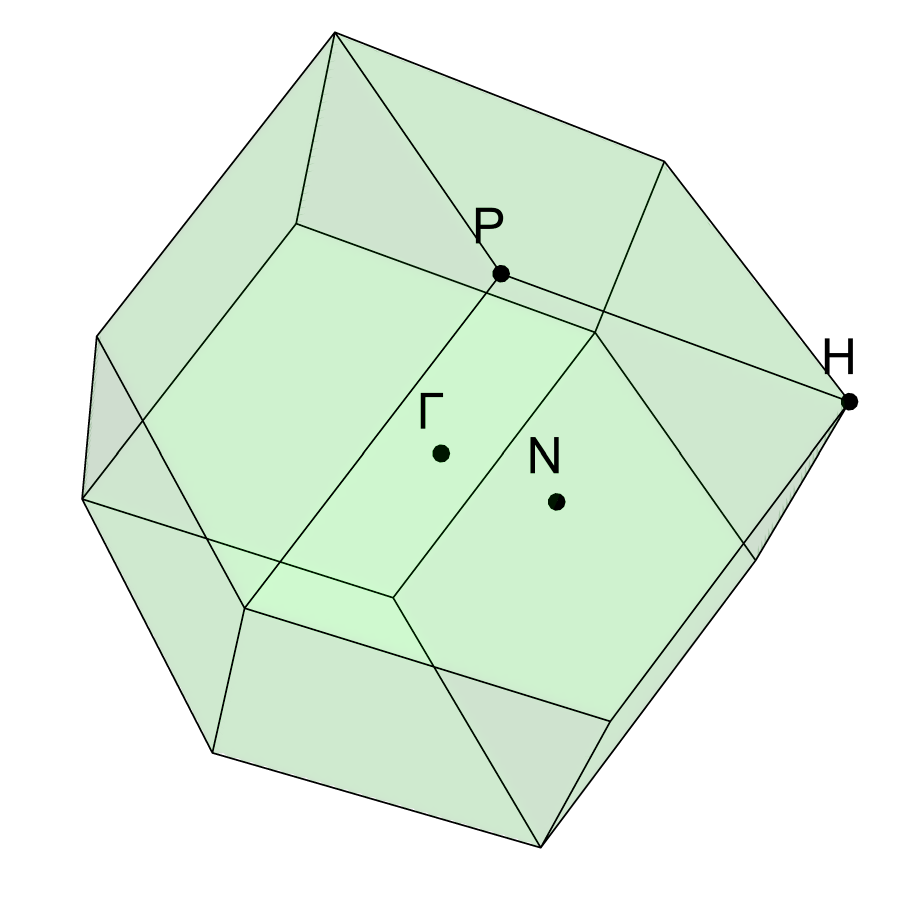}\\ \includegraphics[scale=0.25]{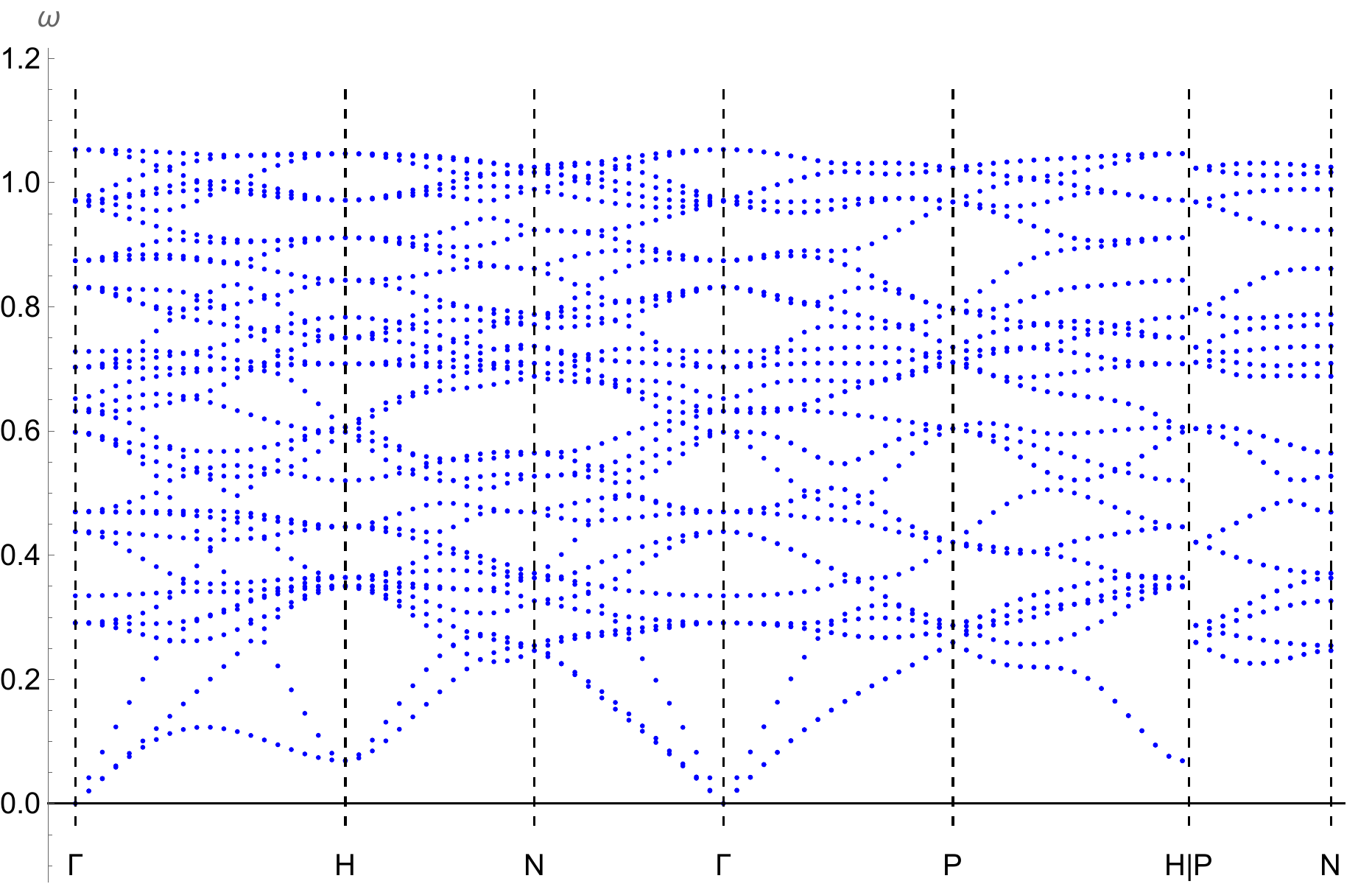}
     \caption{(top left) the structure of the BCC toy model (made using VESTA \cite{vesta}) (top right) its Brillouin zone (bottom) its phonon dispersion}
     \label{fig:cub}
 \end{figure}

 Since the groups $Ibca$, $I4_1/acd$ and $Ia\Bar{3}$ are very similar (they are body centered, nonsymmorphic, and reduce to $Iba2 (45)$ after symmetry breaking), we consider consider only a simplistic model in the group $Ia\Bar{3}$. This group includes materials with the bixbyite structure \cite{bixbyite1,bixbyite2}. Consider a toy model with atoms $A$ at the $8a$ Wyckoff positions, and $B$ at the $16c$ positions, with $B$ having $0.64$ times the mass of $A$. We consider a tight binding model with interactions only between atoms separated by less than $0.5a$ where $a$ is the lattice constant. The model, its Brillouin zone, and its phonon dispersion are illustrated in Fig~\ref{fig:cub}

 In these space groups, the mechanical representations of all Wyckoff positions break down into $4$-fold irreps at the points $P (\pi,\pi,\pi)$. However, they these $4$-fold degeneracies further break into 2 $2$-fold irreps if the crystal symmetry is broken by a polarization along an axis. 

For all three space groups, the tensors $v_{ij}$ and $a_{ij}$ have only diagonal components. For the orthorhombic case, all three components are different; for the cubic case, all three are the same. For the tetragonal case, $v_{xx}=v_{yy}\neq v_{zz}$; $a_{xx}=-a_{yy}$ and $a_{zz}=0$

\begin{figure}
    \centering
    \includegraphics[scale=.15]{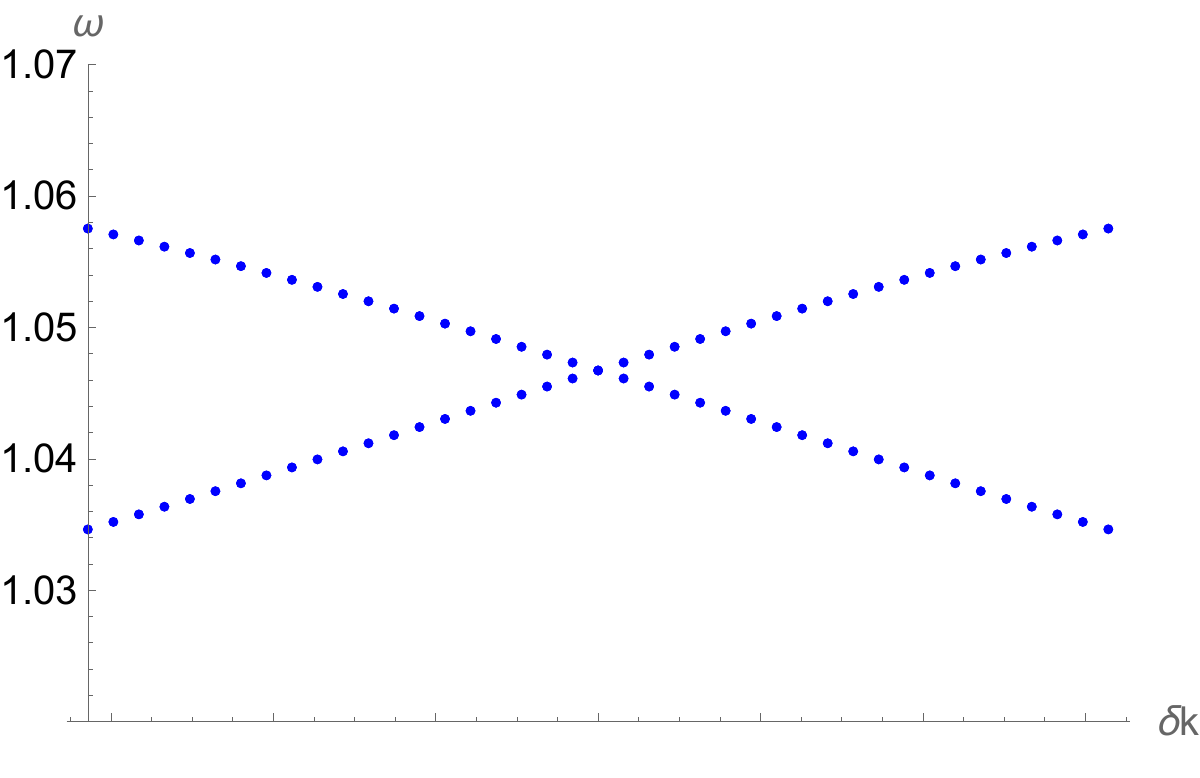}\includegraphics[scale=.15]{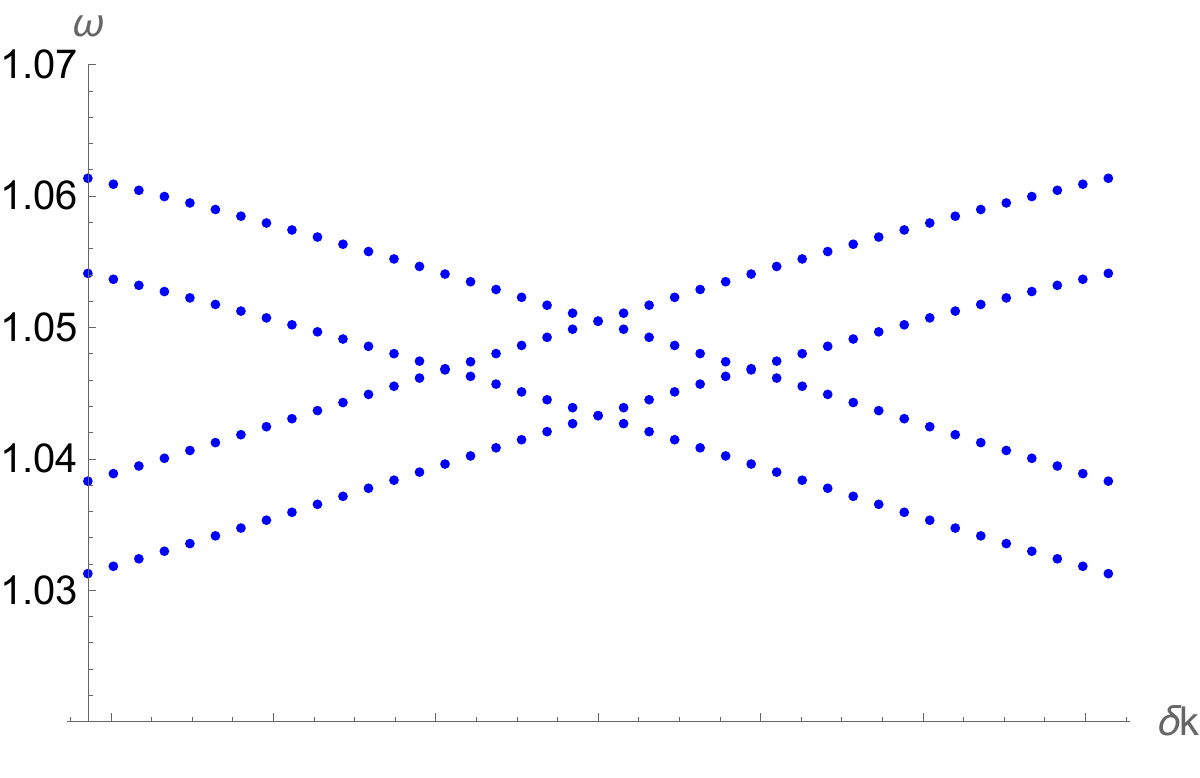}\includegraphics[scale=.15]{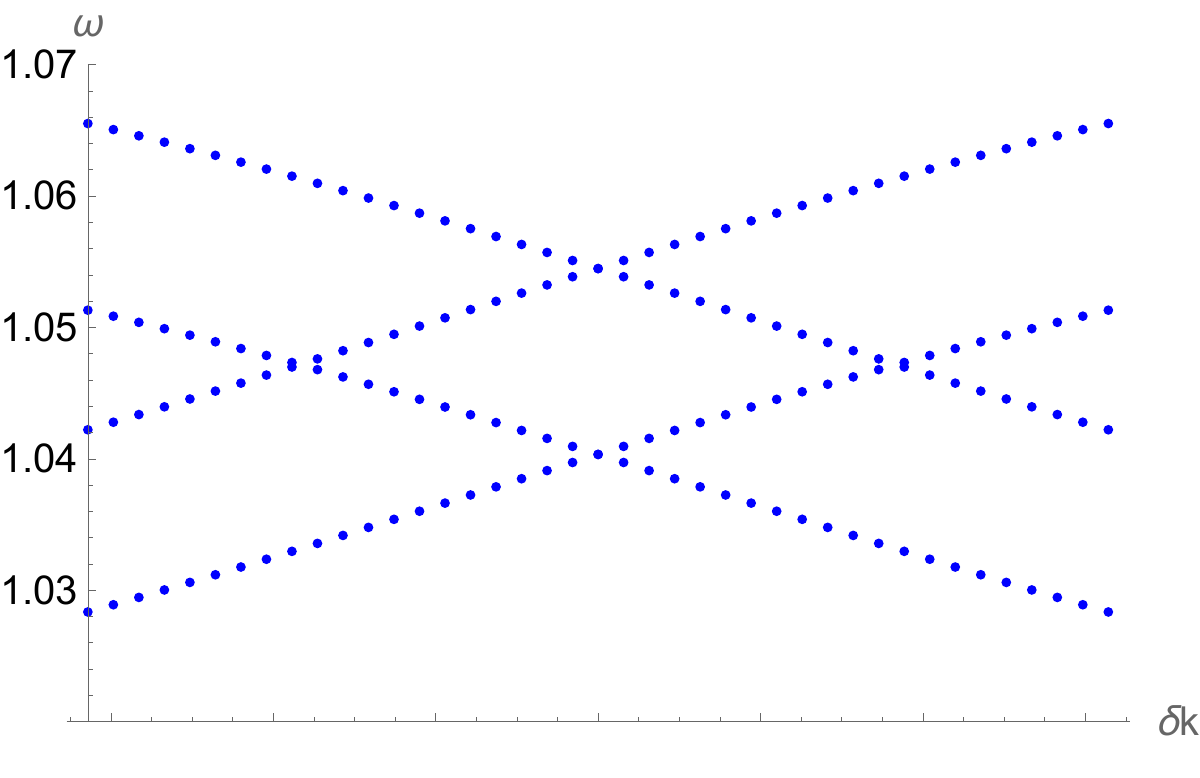}
    \caption{(L-R) The four highest phonon frequencies at $(\pi,\pi,\pi+\delta k)$ vs $\delta k$ for the BCC toy model for $h = 0.00, 0.02, 0.04$ }
    \label{fig:cubplots}
\end{figure}

We consider the case where there is polarization along the $z$ axis. We displace the atoms $A$ by $h$ along $z$, and plot the four highest frequencies for $h = 0.00, 0.02, 0.04$ in Fig~\ref{fig:cubplots}.

\begin{figure}
    \centering
    \includegraphics[scale=0.4]{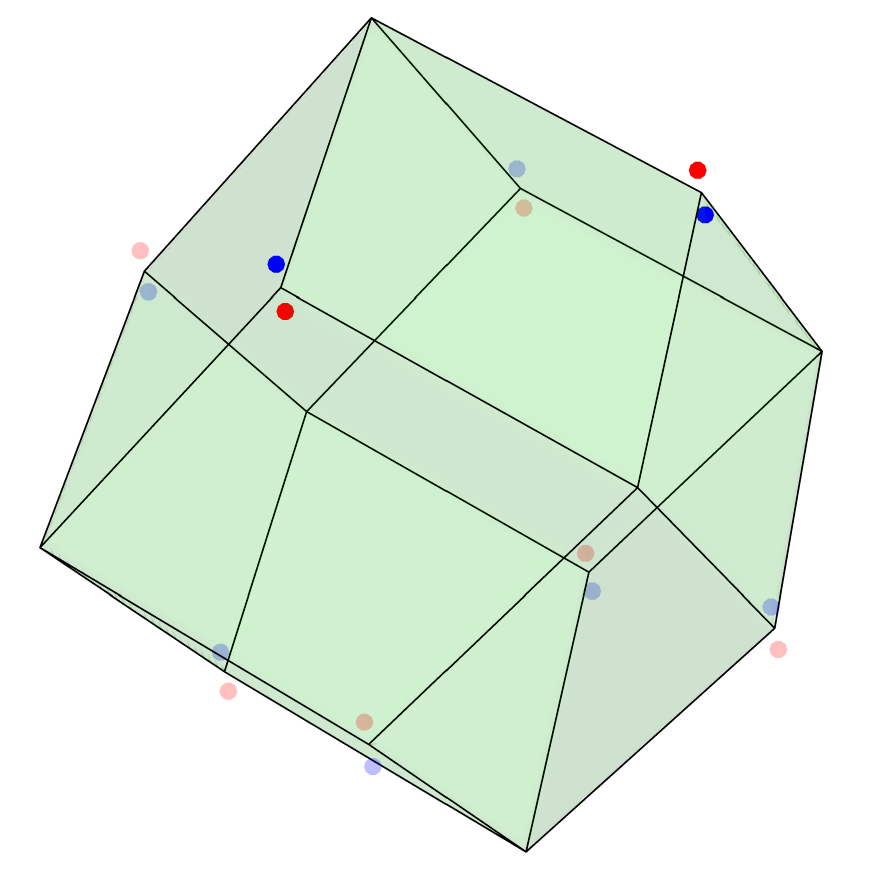}
    \caption{Weyl points in the BCC toy model. There are four distinct points, which are highlighted. The rest are duplicates. Red and Blue indicate opposite chiralities.}
    \label{fig:cubweyls}
\end{figure}

It can be seen that there are Weyl along the $z$ axis. There are 4 points in total, 2 near $P$ and 2 near $-P$, which are illustrated in Fig~\ref{fig:cubweyls}. 

\subsection{Hexagonal}
\begin{figure}
    \centering
    \includegraphics[scale=0.2]{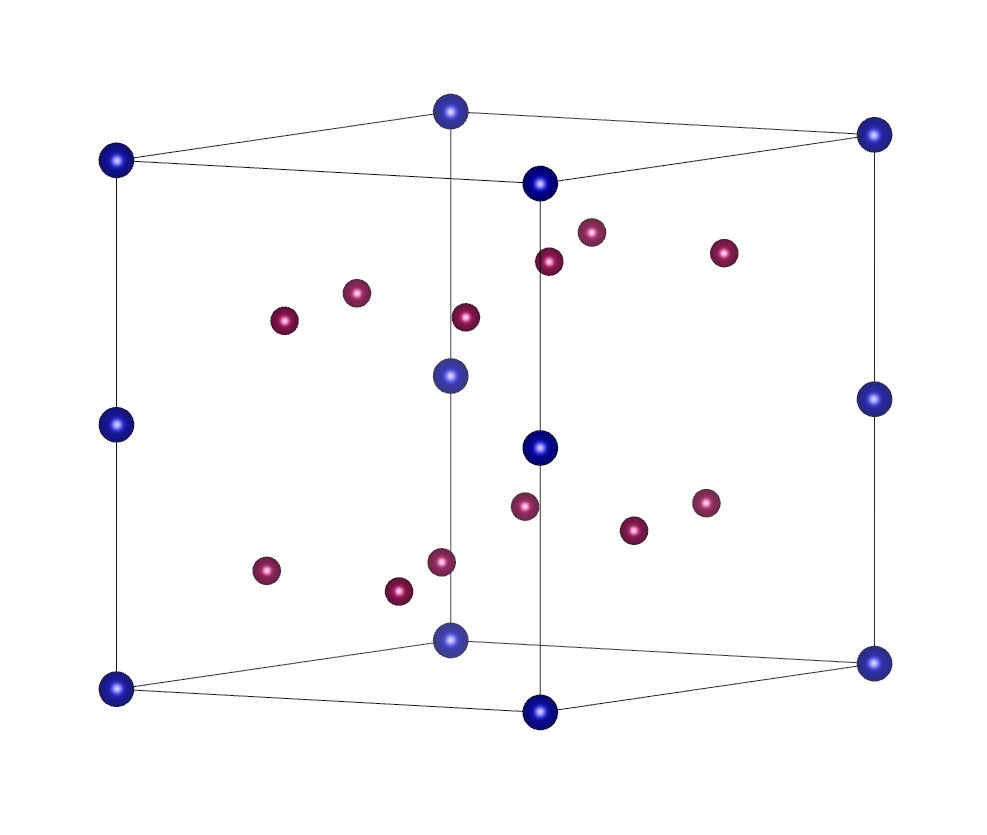}\includegraphics[scale=0.5]{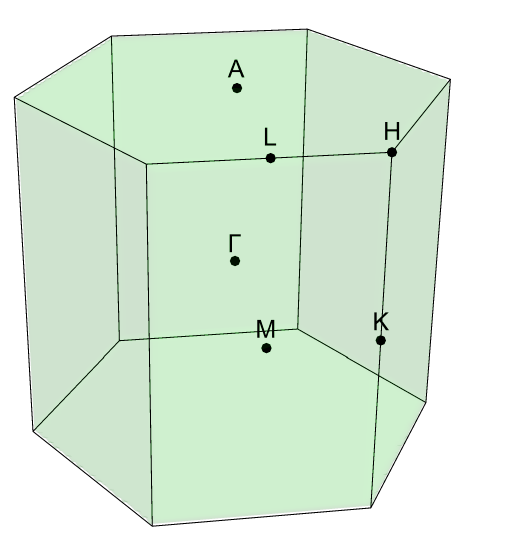}
    \caption{The crystal structure  (made using VESTA \cite{vesta})  and the Brillouin zone of the hexagonal toy model}
    \label{fig:hexmodel}
\end{figure}

Consider the $P\Bar{3}c1$ space group. This space group includes the lanthanide trifluorides \cite{laf}. Consider a toy model with atoms $A$ at the $2a$ and $B$ at the $12g$ positions, with B having $0.64$ times the mass of $A$. The crystal structure and Brillouin zone of this toy model are shown in Fig~\ref{fig:hexmodel}

In this space group, at the point $H (2\pi/3,2\pi/3,\pi)$, each Wyckoff contributes an equal number of $2$-fold and $4$-fold irreps to the mechanical representation. For example, $2a$ contributes $1$ of each irrep, which $12g$ contributes 6. We focus on the $4$-fold irreps.
In this space group, in general, $v_{xx}=v_{yy}\neq v_{zz}$ and $a_{xx}=a_{yy}\neq a_{zz}$

Similar to the BCC case, we shift the positions of atoms $A$ along $z$ by $h=0.004$, and plot how the frequencies change.

\begin{figure}
    \centering
    \includegraphics[scale=0.12]{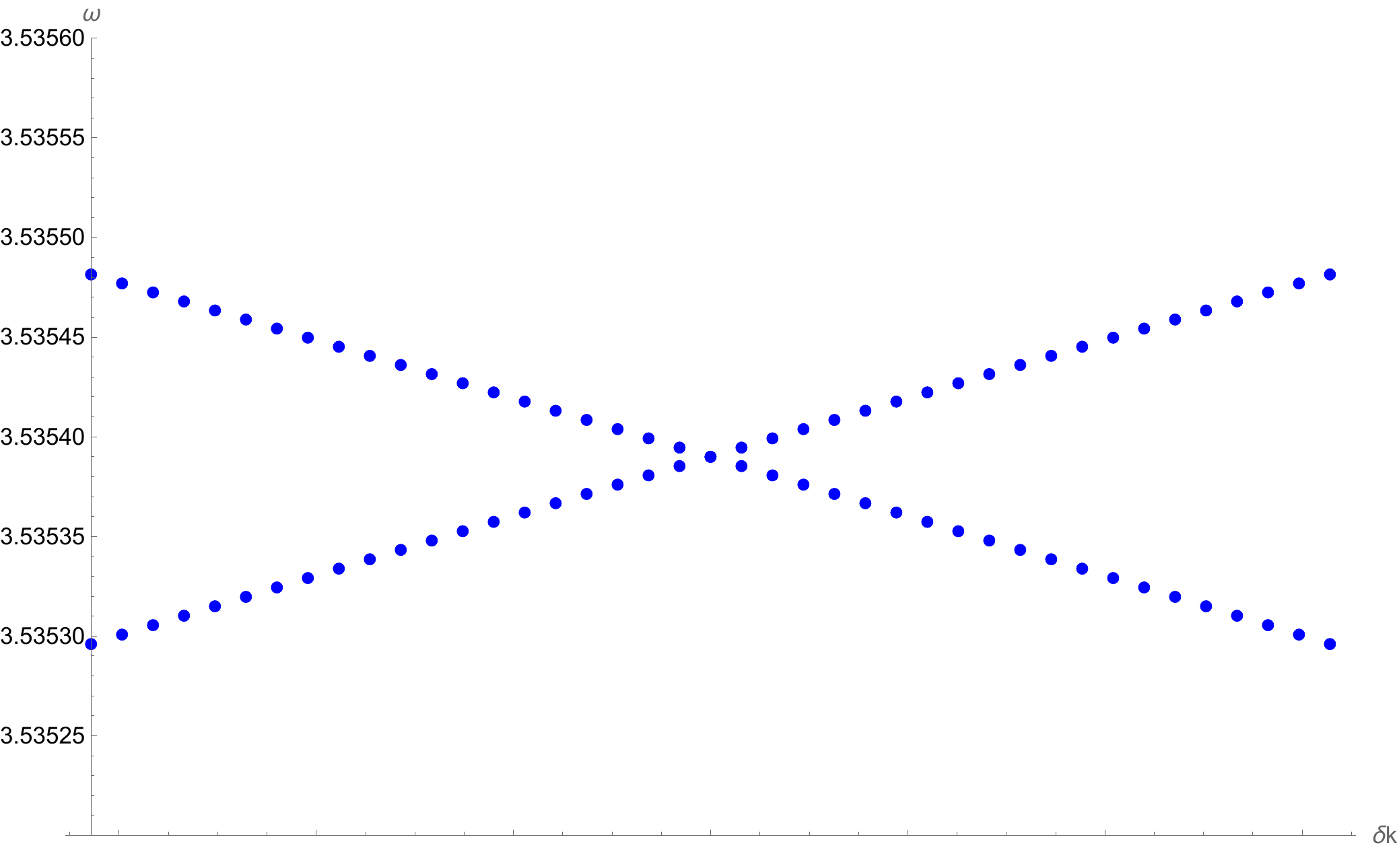}\includegraphics[scale=0.12]{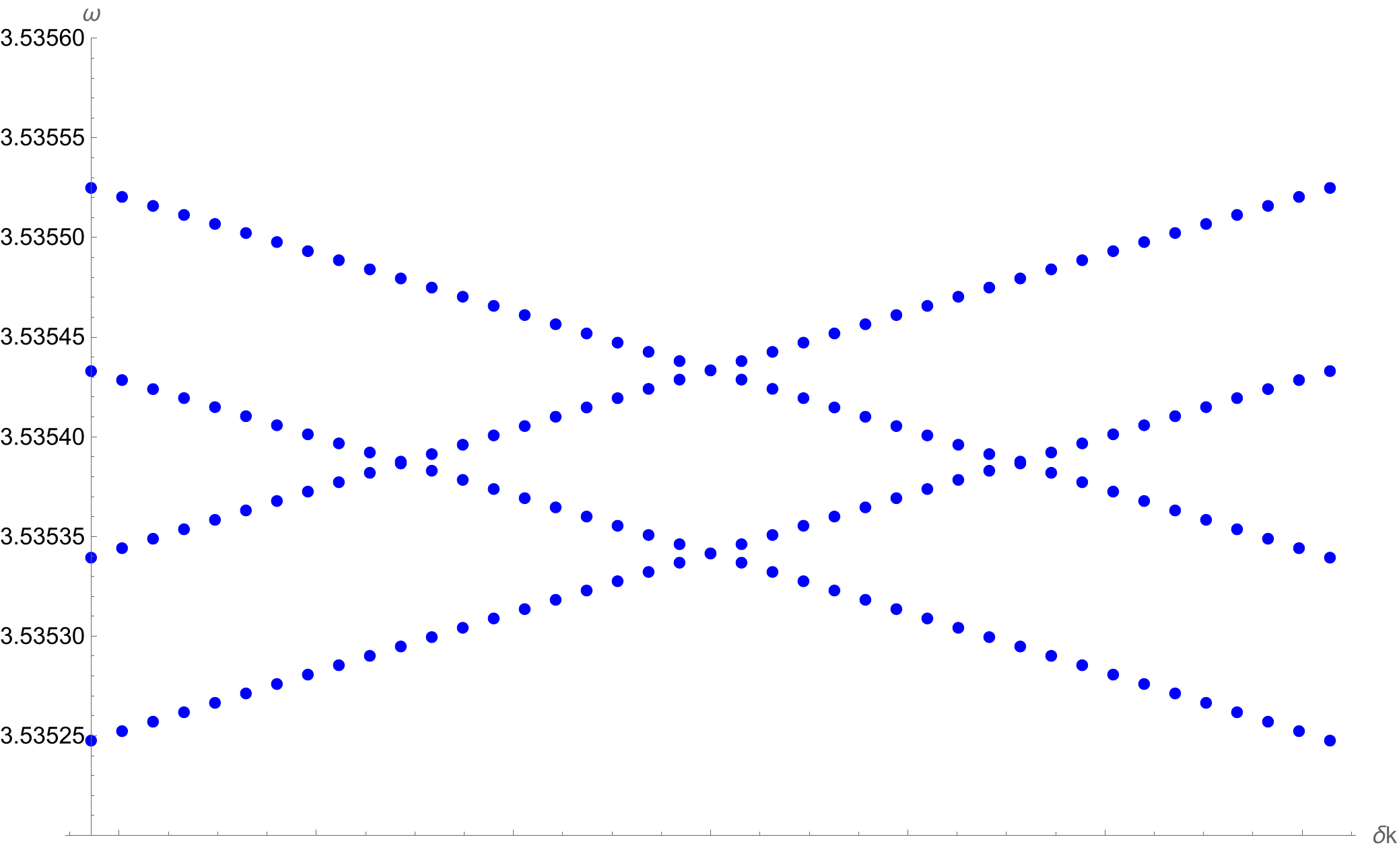}
    \caption{The four highest phonon frequencies at $(2\pi/3,2\pi/3,\pi+\delta k)$ vs $\delta k$ for the hexagonal toy model for $h = 0.000$ (left) and $0.004$ (right) }
    \label{fig:enter-label}
\end{figure}
\begin{figure}[ht]
    \centering
    \includegraphics[scale=0.4]{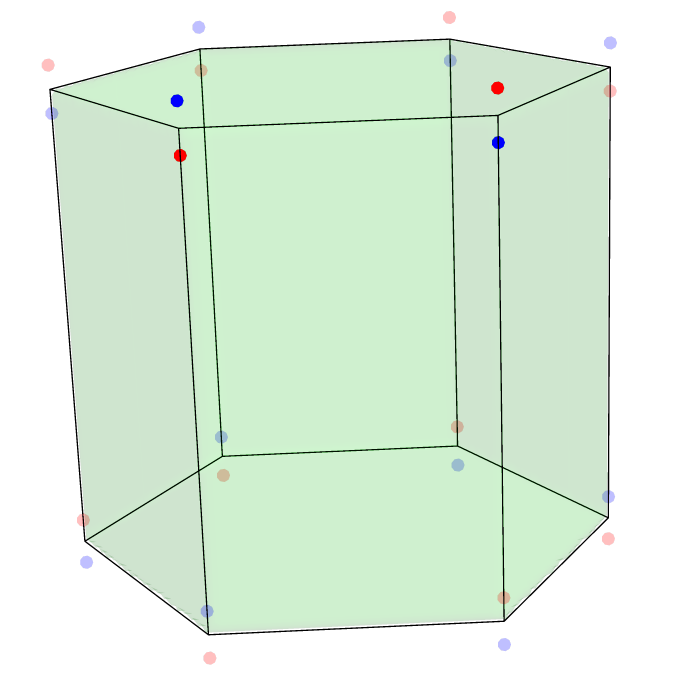}
    \caption{Weyl points in the hexagonal toy model. There are four distinct points, which are highlighted. The rest are duplicates. Red and Blue indicate opposite chiralities.}
    \label{fig:hexweyl}
\end{figure}

Similar to the BCC model, there are $4$ Weyl points, $2$ near $H$ and $2$ near $-H$. They are depicted in Fig~\ref{fig:hexweyl}

 \section{Discussion}
 It has been shown that in four space groups (73, 142, 206, and 165), an electric polarization necessarily produces Weyl phonons. The existence of Weyl phonons depends only on symmetry, and is independent of the crystal structure, lattice constants, and other parameters. The Weyl points could be controlled by changing the polarization. 

 If the projections of Weyl points of opposite chirality onto an external surface do not coincide, that surface will host surface arcs \cite{weyl}. Such surfaces include $(110)$ for the body-centered structures, and $(100)$ for the hexagonal one. As surface phenomena, these arcs would be easier to detect than bulk states.

 Furthermore, if the polarization is due to ferroelectric order, domain walls would host internal surface arcs if the projections of right handed cones in one domain coincide with those of left handed ones in the other domain and vice versa. These internal arcs could be tuned be adjusting the polarization. 

\begin{acknowledgments}
    I acknowledge useful discussions with Egor Babaev, Katia Gallo and Tiantian Zhang.

    NORDITA is funded in part by NordForsk, the Nordic council of ministers.
    \end{acknowledgments}
    \bibliographystyle{apsrev4-2}
\bibliography{refs}

\end{document}